\documentclass[prb,superscriptaddress,twocolumn,showpacs,preprintnumbers,amsmath,amssymb]{revtex4}

\usepackage{graphicx}

\usepackage{dcolumn}
\usepackage{bm}
\usepackage{xspace,longtable,verbatim}
\usepackage{subfig}

\newcommand{\degree}{$^{\circ}$\xspace}
\newcommand{\cmVs}{cm$^2$V$^{-1}$s$^{-1}$\xspace}

\begin{document}
\preprint{Nanotechnology 19 (2008) 485701}

\title{Effect of doping-- and field--induced charge carrier density on the electron transport in nanocrystalline ZnO}

\author{Maria S Hammer}
\email{maria.hammer@physik.uni-wuerzburg.de}
\author{Carsten Deibel}%
\affiliation{Experimental Physics VI, Faculty of Physics, Julius Maximilians University of W\"urzburg, Am Hubland, 97074 W\"urzburg, Germany}

\author{Daniel Rauh}%
\author{Volker Lorrmann}%
\affiliation{Functional Materials for Energy Technology, Bavarian Centre for Applied Energy Research (ZAE Bayern), Am Hubland, 97074 W\"urzburg, Germany}

\author{Vladimir Dyakonov}
\email{vladimir.dyakonov@physik.uni-wuerzburg.de}
\affiliation{Experimental Physics VI, Faculty of Physics, Julius Maximilians University of W\"urzburg, Am Hubland, 97074 W\"urzburg, Germany}
\affiliation{Functional Materials for Energy Technology, Bavarian Centre for Applied Energy Research (ZAE Bayern), Am Hubland, 97074 W\"urzburg, Germany}

\date{\today}
\begin{abstract}

Charge transport properties of thin films of sol--gel processed undoped and Al-doped zinc oxide nanoparticles with variable doping level between 0.8~at$\%$ and 10~at$\%$ were investigated. The X-ray diffraction studies revealed a decrease of the average crystallite sizes in highly doped samples. We provide estimates of the conductivity and the resulting charge carrier densities with respect to the doping level. The increase of charge carrier density due to extrinsic doping were compared to the accumulation of charge carriers in field effect transistor structures. This allowed to assess the scattering effects due to extrinsic doping on the electron mobility. The latter decreases from 4.6$\cdot$10$^{-3}$~\cmVs to 4.5$\cdot$10$^{-4}$~\cmVs with increasing doping density. In contrast, the accumulation leads to an increasing mobility up to 1.5$\cdot$10$^{-2}$~\cmVs. The potential barrier heights related to grain boundaries between the crystallites were derived from temperature dependent mobility measurements. The extrinsic doping initially leads to a grain boundary barrier height lowering, followed by an increase due to doping-induced structural defects. We conclude that the conductivity of sol--gel processed nanocrystalline ZnO:Al is governed by an interplay of the enhanced charge carrier density and the doping-induced charge carrier scattering effects, achieving a maximum at 0.8~at$\%$ in our case.   
\end{abstract}

\vspace{2pc}
\maketitle

\section{Introduction}
Semiconducting zinc oxide nanocrystallites (nc-ZnO) and their optical, electrical and structural properties have been thoroughly investigated in recent years.\cite{orlinskii2008, xue2006, lin2008, paul2002, schuler1999, sagar2005, alam2001} The ability to synthesize nc-ZnO via the sol--gel technique makes them an interesting choice for versatile applications in solution processable electronics such as transparent transistors~\cite{cheng2007} or hybrid organic--inorganic solar cells~\cite{beek2004, gilot2007a, gilot2007b, hadipour2008}. In photovoltaics, the nc-ZnO was successfully employed to improve the charge generation in donor--acceptor blends.~\cite{beek2004, koster2007}

The bulk properties of zinc oxide, a II-VI semiconductor, are well known. It has a high dielectric constant of around 8,~\cite{langton1958}which is advantageous, e.g., for the charge separation process in hybrid solar cells, a high electron affinity,~\cite{marien1976} a large band gap of 3.3 eV~\cite{srikant1998} and p- and n-dopability. In contrast, the physical properties as well as the technological issues of zinc oxide nanoparticles remain to be clarified in view of several aspects. In Ref.~\cite{zhong2004}, technological concepts including  the sol--gel preparation route are reviewed. There are several routes to synthesize Al-doped zinc oxide nanoparticles.~\cite{ xue2006, lin2008, sagar2005, schuler1999, meulenkamp1998} In sol--gel synthesized nc-ZnO:Al, the Zn$^{2+}$ ions are substituted by  Al$^{3+}$ ions, as has been recently proven by Orlinskii $et$ $al.$\cite{orlinskii2008}
The intention of extrinsic doping of ZnO nanoparticles is to have a well-defined tool to control the electrical conductivity by generating various free charge carrier densities and to keep the advantages of the processability at the same time. 

Aluminium acts as an electron donor in ZnO and introduces an allowed energy state just below the conduction band energy. The thermal release of electrons into the band leads to an increase in charge carrier density and, consequently, conductivity. The electrical transport properties of ZnO are mainly investigated by four probe conductivity technique as well as by Hall effect measurements.~\cite{lin2008, paul2002, sagar2005, schuler1999, alam2001} In crystalline systems, doping usually lowers the mobility due to the enhanced ionic scattering.~\cite{ellmer2008} In contrast, a positive effect of doping on the mobility in polycrystalline and nanocrystalline ZnO systems at high doping levels has been reported. \cite{orton1980, seto1975, baccarani1978} Doping has also a strong influence on the crystallite size.\cite{paul2002, schuler1999} The simulations of charge transport within field effect transistors (FET) performed in Ref.~\cite{hossain2003} ascertained the influence of the grain boundary density on charge carrier mobility.

In this work, the structural and electrical transport properties of the sol--gel processed ZnO nanoparticles  with variable Al-doping level were investigated by using X-ray diffraction (XRD), scanning electron microscopy (SEM) and field-effect transistor (FET) measurements. The use of FET provides a tool to determine the mobility, the conductivity and the charge carrier density. Its advantage lies in the variation of the charge carrier density at a constant defect density. We compare the influence of  charge carrier density achieved by extrinsic substitutional doping and by the gate-induced accumulation of charges on the mobility. The doping-induced crystallite size variation and its consequences for the charge transport are discussed.

\section{Experimental}
\subsection{Materials and Synthesis}
The synthesis of nc-ZnO was performed according to the route introduced by Xue $et$ $al.$~\cite{xue2006} It offers the advantage of the functionalization of the particles by using organic groups. These organic compounds act as a surface coating of the particles and help to suppress agglomeration within the solution. 
2-methoxyethanole (MTE), monoethanolamine (MEA), zinkacetate dihydrate (ZAD) and aluminiumnitrate nonahydrate (ANN) were purchased from Sigma-Aldrich. 2.096 ml of the stabilizer MEA were dissolved in 100~ml of the solvent MTE to lead to a 0.35~M solution. 0.035~mol of the starting material ZAD was added. ANN depending on the designated doping level was admixed to the solution to achieve up to 10~at$\%$ Al doping. The solution was stirred for 2~h at 60\degree{}C and subsequently aged for one day.

\subsection{Sample Preparation and Measurement}
The field effect transistors were built upon Sb--doped silicon wafers purchased at Si-Mat. The highly n-doped Si (001) with a nominal resistivity ranging from 0.01~$\Omega$m to 0.02~$\Omega$m serves as the gate electrode. The top layer of the wafer consists of 200~nm thermally grown SiO$_2$ which forms the dielectric. We used photolithography to get well defined source--drain structures. We thermally evaporated 1~nm Ti as an adhesion layer and subsequently 19~nm thick Au contacts on top of the dielectric.
After cleaning the substrate, the nanoparticle solution was spin coated for one minute at 3000~rpm. Thereafter the substrate was heated to 300\degree{}C in an oven in ambient atmosphere for one hour. This step is necessary to remove MEA. Subsequently, the samples were sintered at 500\degree{}C in order to improve the particle--particle cohesion. The layer thickness is about 25~nm and does not significantly vary with doping level. It was determined using a Profilometer (Dektak 150).

The current--voltage measurements were carried out using an Agilent Parameter Analyzer 4155C. The samples were located in a closed cycle cryostat under vacuum conditions and in the dark for at least 24 hours before the measurements were carried out at 300~K, if not specified otherwise.

Scanning Electron Microscope (SEM) micrographs were made using a Zeiss ``ultra~plus'' SEM. Here, Si substrates were used.

The X-ray diffraction (XRD) using Cu K$_{\alpha}$ was carried out on glass substrates, as the Si/SiO$_2$ substrates show X-ray diffraction peaks near those of ZnO in Wurtzite structure. In order to get a good signal to noise ratio, thicker nc:ZnO:Al layers as in the SEM and FET studies were used. This was achieved by applying the nanoparticle solution five times. The sample was heated to 300\degree{}C after each spin coating step to remove MEA from the layer and, hence, to avoid carbon compounds encapsulated in the nanoparticle pores. Subsequently, the sintering step at 500\degree{}C was applied. The thickness of the layer is about 120~nm and it is almost independent of the doping level.

\subsection{Structural analysis}
\begin{figure} 
   \includegraphics{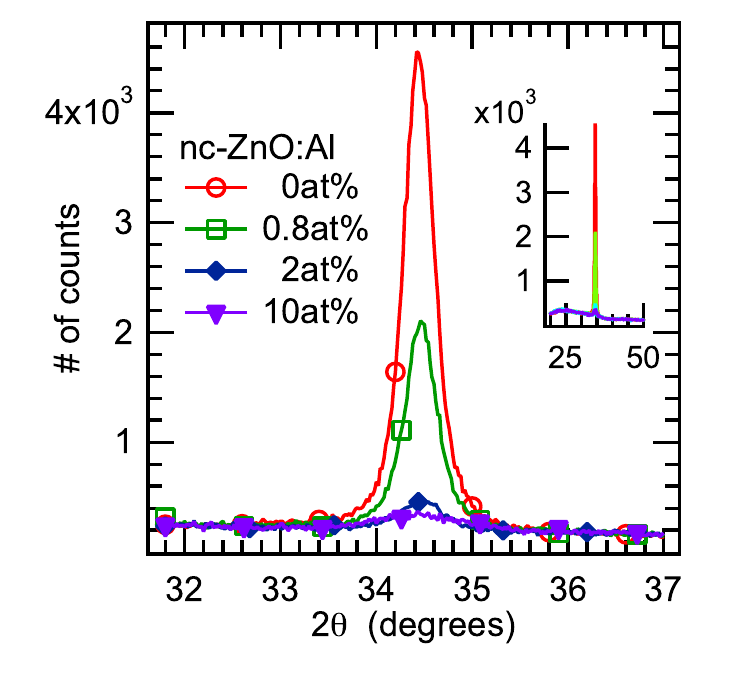} 
   \caption{(Color online) X-ray diffraction of ZnO nanoparticles with varying Al doping level as indicated by the legend. Inset: The full width of $\theta$ indicates  only one peak corresponding to the (002) direction of the ZnO Wurtzite structure.}
   \label{fig:xrd}
\end{figure}

\begin{table}
	\caption{Average crystallite sizes of undoped and variedly Al--doped nc-ZnO estimated from XRD spectra using the Debye-Scherrer formula.}
	\begin{center}
		\begin{tabular}{|c|c|c|c|c|}\hline  & 0~at$\%$ & 0.8~at$\%$ & 2~at$\%$ & 10~at$\%$ 		\\\hline size/nm & 20.8 & 19.6 & 15.0 & 7.2 \\\hline 
		\end{tabular}
	\end{center}
	\label{dbs}
\end{table}%

In order to determine the structure and orientation of the nanocrystals we carried out XRD measurements.
The XRD spectra (Figure \ref{fig:xrd}) show only one peak corresponding to the orientation along the c-axis ((002) peak) of the Wurtzite structure. Other peaks resulting from different orientations of the nanoparticles on the substrate are not observable. This indicates that the particles are highly oriented on the substrate. Depending on the doping level, the maximum of the (002) peak decreases while the peak width broadens. An estimation of the nanocrystallite size was done by using the Debye-Scherrer formula $\beta$$_{1/2}$ = (0.94$\lambda$)/($l$$\cdot$cos$\theta$), where $l$ is the particle diameter, $\lambda$ the X-ray wavelength, $\beta$$_{1/2}$ the full width at half maximum of the peak and $\theta$ the peak maximum position.\cite{jones1938} A shrinking from about 20~nm in the undoped case to about 7~nm for the highest doping level is found, as seen in Table \ref{dbs}. Although the estimation of the particle size at the highest doping level is less accurate due to the low peak intensity and the large peak width, the tendency of a decreasing crystallite size with increasing extrinsic doping level is clearly visible. 
\begin{figure}
  	 \includegraphics{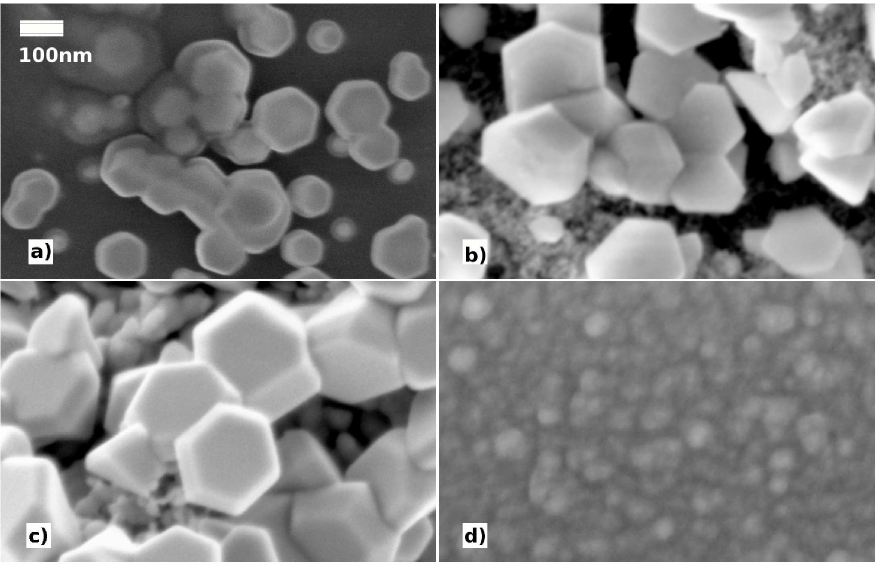} 
  	 \caption{SEM micrographs taken with 3~kV acceleration voltage and using the inlens detector. The images give a clear indication of the hexagonal structure and c-axis orientation of the undoped and variedly doped nc-ZnO: a) 0~at$\%$; b) 0.8~at$\%$; c) 2~at$\%$, d) 10~at$\%$. Highly n-doped Si/SiO$_2$ was used as substrate.}
   	\label{fig:sem}
\end{figure}

The XRD spectra give evidence of the average crystallite size, but the sintering induced linking between particles is important to get continuous electron pathways for the charge transport. Note that one particle can consist of several crystallites. Indeed, the sintering was successfull for 0~at$\%$, 0.8~at$\%$,  2~at$\%$ and to a lesser extent for 10~at$\%$ doping, as demonstrated by the SEM micrographs in Figure~\ref{fig:sem}. The particles also exhibit a hexagonal geometry, which is a clear indication of  the c-axis orientation. The SEM micrographs show only the dominant particle sizes in a restricted region of the sample, and therefore do not necessarily represent the average crystallite size extracted from the XRD measurements.

\subsection{Field effect transistor measurements} \label{fet0at}

\begin{figure}
	\subfloat[]{\includegraphics{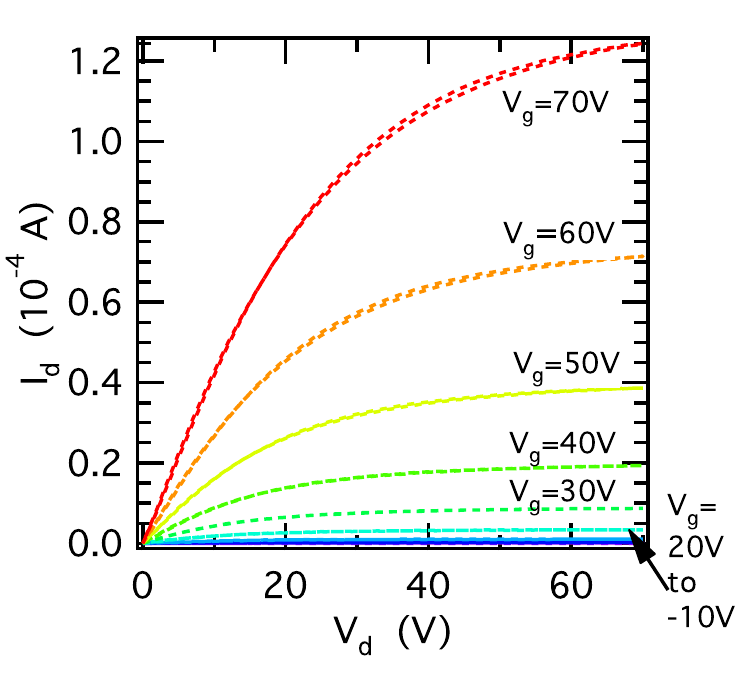}\label{fig:0atouta}}%
	\qquad
	\subfloat[]{\includegraphics{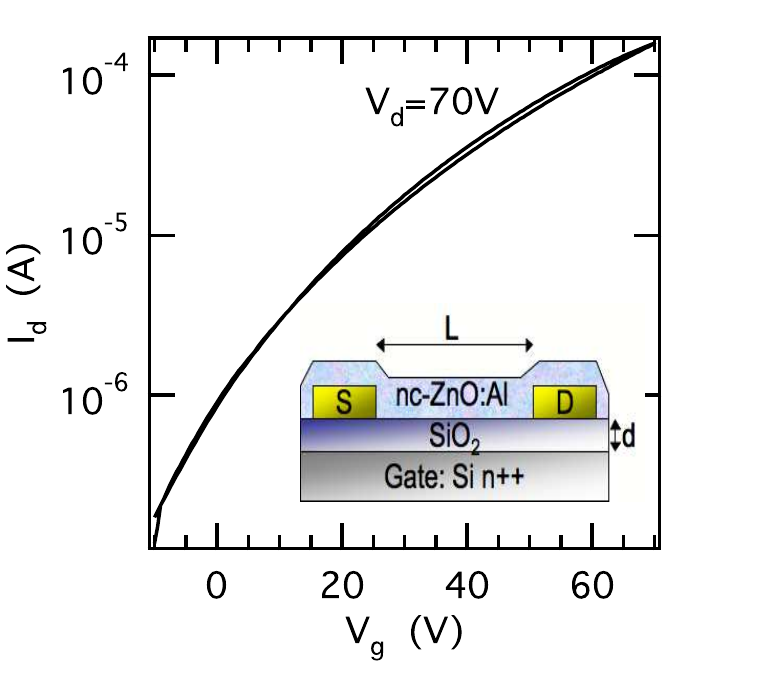}\label{fig:0atoutb}}%
	\caption{(Color online) Field effect transistor characteristics of undoped zinc oxide nanoparticles at T~=~300~K: \ref{fig:0atouta}) Output characteristic: $I_d$--$V_d$ in dependence of $V_g$ as indicated by the legend \ref{fig:0atoutb}.) Transfer characteristic $I_d$--$V_g$ at $V_d$=70~V and schematic of the FET device (S: source, D: drain, L: channel length, d: thickness of the dielectric). Transistor details: $L$=160~$\mu$m, Au bottom contacts on SiO$_2$ dielectric.}
   	\label{fig:0atout}
\end{figure}

For the investigation of the charge transport properties we chose the field effect transistor geometry.  It provides a way to determine the charge carrier mobility for various carrier densities without affecting the defect density. A sketch of the used device structure is shown in Figure \ref{fig:0atout}.

The transistor characteristics of the nominally undoped sample (Figure \ref{fig:0atout}) indicate the effect of  accumulating charges by applying a gate voltage. The output characteristics (Figure \ref{fig:0atouta}) show the drain current $I_d$ $vs.$ drain voltage $V_d$ for several constant gate voltages $V_g$. The output characteristics indicate the transistor performance for the n-conducting case at positive $V_g$. There is no significant injection barrier at the gold/nc--ZnO interface, as can be seen from the ohmic behavior at small $V_d$.

The transfer characteristics $I_d$ $vs.$ $V_g$ at a constant source--drain field of 70~V (Figure \ref{fig:0atoutb}) provide a way to determine the charge carrier mobility. The accumulation of charges close to the dielectric due to the applied gate voltage modifies the charge carrier density and, hence, the conductivity in the channel. Accordingly, the drain current increases with raising gate voltage. The mobility in this case is proportional to the derivative of the $I_d$--$V_g$ curve.

The depletion of the charge carriers in the channel due to decreasing gate voltage usually results in a constant low level drain current, the off-current. In the n-conducting case, one estimates the off-current at low gate voltages. In the present study, a rather high off-current was observed, pointing at a high charge carrier concentration being initially present in the channel as a result of unintentional doping.

In the following, we focus on the effect of extrinsic doping on the charge transport, i.e., the conductivity, free charge carrier density, mobility and also the barrier height at the nanocrystallite interface.

\subsubsection{Conductivity and charge carrier density}

\begin{figure} 
   \includegraphics{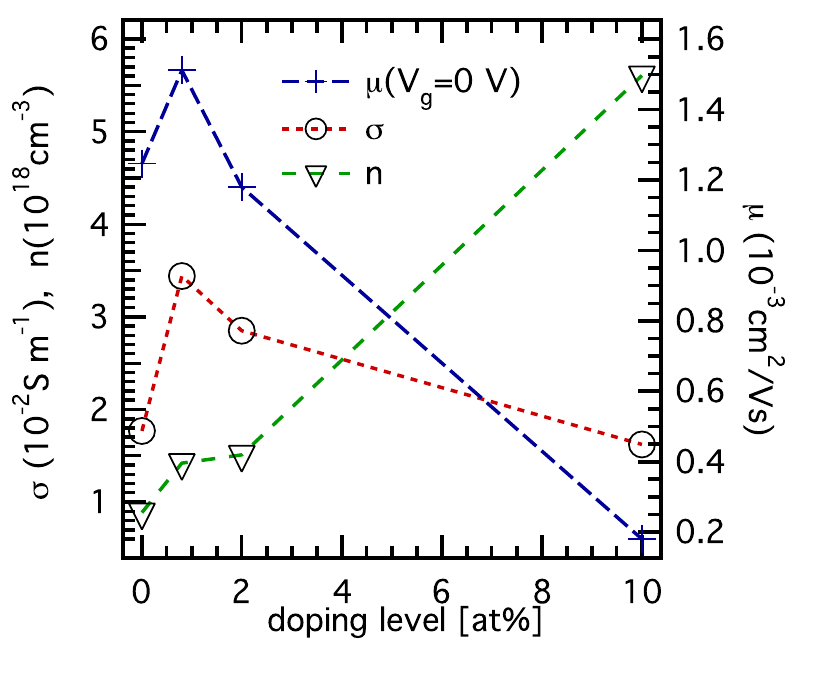} 
   \caption{(Color online) Conductivity (circles), free charge carrier density (triangles) and mobility (crosses) $vs.$ doping level of zinc oxide nanoparticles at T=300~K; dashed lines: guides to the eye. Sample details: Au bottom contacts on SiO$_2$ dielectric.}
   \label{fig:mucondfitn}
\end{figure}

The conductivity $\sigma$, the free charge carrier density $n$ and the mobility $\mu$ are related by
\begin{equation}
\sigma=q n \mu,
\label{sigma}
\end{equation} 
where $q$ is the elementary charge. Extrinsic doping is supposed to cause thermally activated free charge carriers and, accordingly, a higher conductivity is expected.

In order to estimate the values of $\sigma$ and $n$, we measured $I_d$--$V_d$ characteristics at zero gate voltage in the linear regime, i.e. at low drain voltages, and extracted $\sigma$ using Ohm's law $\sigma = j/(V_d/L)$, where $j$ is the current density. $j$ results from $I_d$ normalized for the contact area, which is the product of the channel width $W$ and the minimum height of the transport layer or the injecting electrode, amounting to 20~nm in our case. $L$ is the channel length of the transistor. Figure \ref{fig:mucondfitn} shows the values of $\sigma$ (circles) for the undoped and the variously Al-doped nc-ZnO. The conductivity of the nominally undoped nc-ZnO was found to be 1.8$\cdot$10$^{-2}$~Sm$^{-1}$. $\sigma$ reaches a maximum of 3.4$\cdot$ 10$^{-2}$~Sm$^{-1}$ at 0.8~at$\%$ doping level. The high doping level corresponds to 1.8$\cdot$10$^{-2}$~Sm$^{-1}$ similar to the initial value of the nominally undoped material. 

The field effect mobility (indicated by the crosses in Figure~\ref{fig:mucondfitn}), which will be addressed in detail in the next subsection, and the conductivity have been used to calculate the free charge carrier density using equation (\ref{sigma}). The extracted values for $n$ are shown in Figure~\ref{fig:mucondfitn} (triangles). As expected, the free charge carrier density increases with the extrinsic doping concentration from 8.9$\cdot$10$^{17}$~cm$^{-3}$ to 5.6$\cdot$10$^{18}$~cm$^{-3}$.

\subsubsection{Mobility}

\begin{figure} 
   	\includegraphics{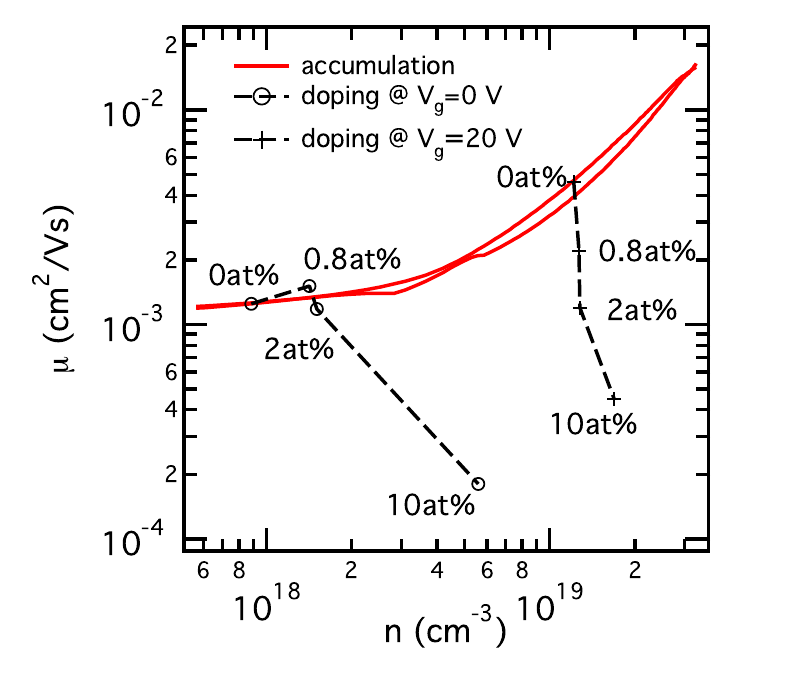} 
   	\caption{Mobility $vs.$ charge carrier density at T=300~K. Solid line: increasing mobility due to accumulation within the field effect channel. Symbols: FET mobility $vs.$ doping: Circles correspond to $V_g$=0~V, crosses to $V_g$=20~V, dashed lines: guides to the eye. Note: forward and backward accumulation directions show little charging effects. The variation of the entire charge carrier density (see text) due to doping is smaller than due to the field effect accumulation.}
   	\label{fig:accdope}
\end{figure}

Here we focus on the effect of free charge carrier density on the mobility, due to both, accumulation as well as  extrinsic doping.

We chose a channel length of $L$=160~$\mu$m since in this case the channel resistance exceeds the contact resistance. Consequently, the linear and saturation mobility are almost identical (not shown). 
The mobility data in this work was derived from the saturation regime, where the drain current is given by
\begin{equation}
	I_d=\mu(W/2L)C_i(V_g-V_t)^2,
\end{equation}
where $C_i$ is the capacitance per unit area. As it was not possible to determine the off-value of the transistor, the threshold voltage $V_t$ could not be extracted. We therefore neglected $V_t$, noting that this can imply a misestimation of the intrinsic and doping-induced charge carrier density, respectively. However, the general dependence of the transport parameters remains unchanged as the doping-induced variation in $n$, which directly contributes to $V_t$, is on a much smaller scale than the variation by accumulation (Figure \ref{fig:accdope}).

In order to estimate a three dimensional charge carrier density within the channel, we assumed that accumulation takes place within the first two nanometers, i.e. $d_{acc}$=2~nm. The charge carrier density resulting from the field effect measurement, $n_{acc}$, can then be determined from $n_{acc}=C_i (qd_{acc})^{-1}V_g$. 

We note that the assumed accumulation layer of 2~nm is a rough estimate. It has been reported that the accumulation layer depends on the doping level (see, for example, Ref.~\cite{shur1989}). Since the charge carrier density varies only on a small scale due to doping, as compared to the variation due to accumulation, we do not expect this effect to be high. Furthermore, we accounted for additional free charge carriers, as we calculate a total free charge carrier density, which has its origin in doping \textit{and} accumumulation, i.e., the initial intrinsic and extrinsic charge carrier densities were added to get the entire free charge carrier density in the channel.

The mobility $vs.$ charge carrier density plot (Figure \ref{fig:accdope}) indicates that accumulation of charges within the undoped nc-ZnO results in an electron mobility increase over one order of magnitude from 1.3$\cdot$10$^{-3}$~\cmVs to 1.5$\cdot$10$^{-2}$~\cmVs. This is consistent with Hossain $et$ $al.$ ~\cite{hossain2003} who simulated grain boundary barrier modulation within ZnO FETs and concluded that an increasing gate voltage leads to a barrier height lowering.

Since doping leads to an increased charge carrier density, a higher mobility with increasing doping level similar to the effect of accumulation is expected.
For clarity, we present the field effect mobility at various doping levels at $V_g$~=~0~V and $V_g$~=~20~V (Figure \ref{fig:accdope}). The variation of the mobility at $V_g$~=~0~V is also indicated by the circles in Figure~\ref{fig:mucondfitn}.
For $V_g$=0~V, the charge carrier mobility is slightly higher at 0.8~at$\%$ as for the undoped case. 
In contrast, at $V_g$=20~V, $\mu$ decreases monotonously with raising extrinsic doping level from 4.6$\cdot$10$^{-3}$~\cmVs to 4.5$\cdot$10$^{-4}$~\cmVs. The decrease is attributed to doping-induced disorder and dopants generated ionized impurities, acting as scattering centers. We mainly attribute the origin of the mobility lowering to the grain boundary density within the transistor channel. The simulations in Ref.~\cite{hossain2003} indicate that the mobility is exponentially lowered with the number of grain boundaries within the transistor channel and therefore with decreasing particle size. Our experimental data indeed reveal an exponential relation between the mobility at $V_g$=20~V and the doping level (not shown). As the mobility is not only determined by the crystallite size and disorder, but also scattering effects, such as electron--electron and electron--ion scattering, it is not possible to give an explicit dependence between the crystallite size and the mobility.

\subsubsection{Potential barriers at the crystallite boundaries}

\begin{figure}
   	\includegraphics{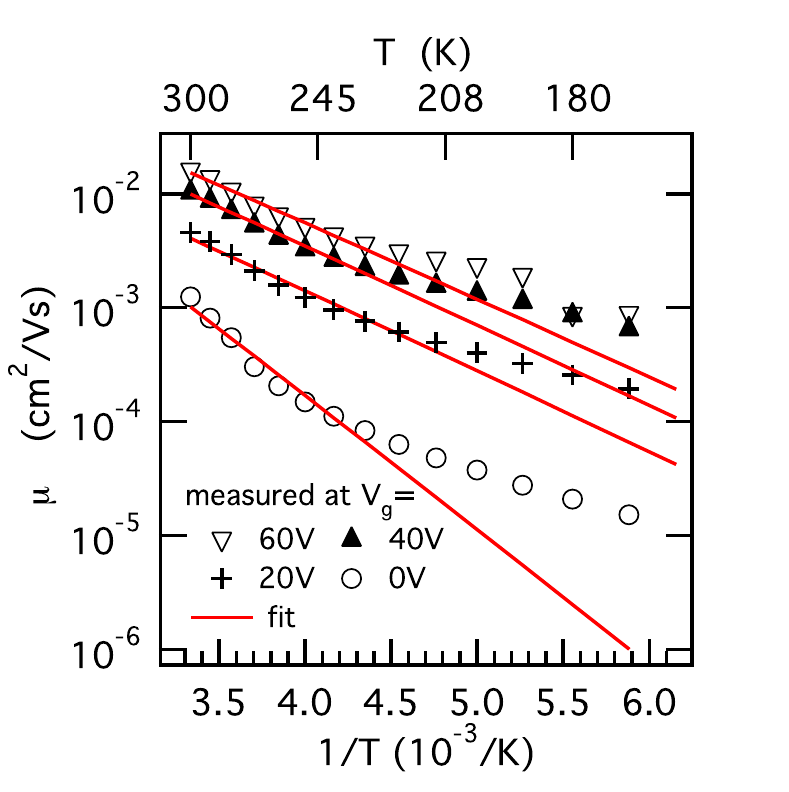} 
   	\caption{(Color online) Temperature dependent mobility data of undoped nc--ZnO (crosses) taken at various $V_g$ as indicated by the legend. The solid lines represent fits according to Eqn.~(\ref{muT}). Two different transport regimes at high and low temperatures can be distinguished.}
   	\label{fig:mut0at}
\end{figure}

\begin{figure} 
   \includegraphics{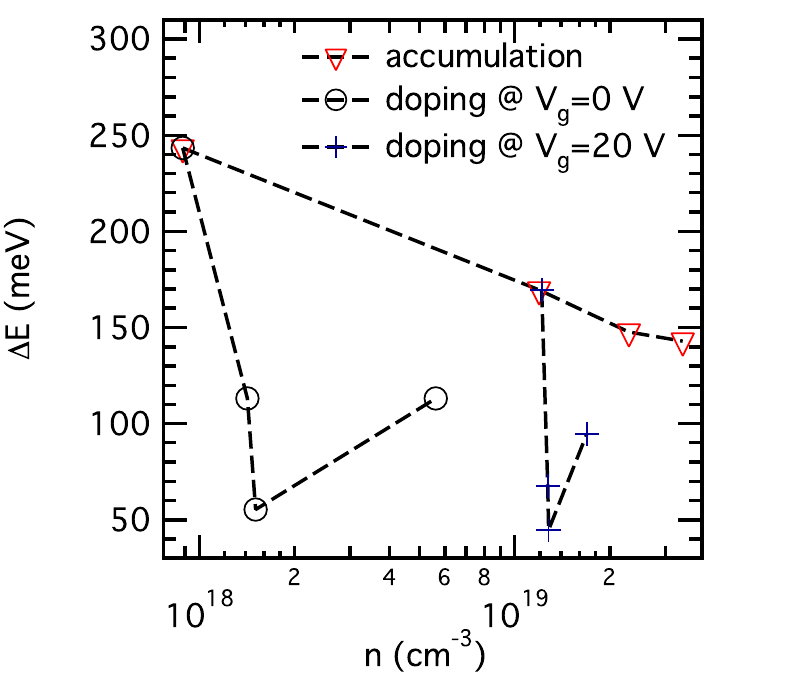} 
   \caption{Potential barrier height $\Delta$$E$  $vs.$ charge carrier density derived from temperature dependent mobility data at various $V_g$ using Eqn.~(\ref{muT}). Comparison of the effect of doping (circles) and accumulation (triangles). The crosses represent the barrier height for various doping levels at $V_g$=20~V. Dashed lines: guides to the eye}
   \label{fig:deltaE}
\end{figure}

In order to investigate the charge transport in more detail, we performed temperature dependent mobility measurements. Figure~\ref{fig:mut0at} exemplarily shows the mobility of undoped nc-ZnO at various $V_g$. The mobility of the nanocrystalline material consists of the contributions from the bulk of the crystallites and from the grain boundaries. If thermally activated hopping over the energy barriers between the crystallites is the dominant transport limiting factor in nc-ZnO,~\cite{kimura2001} the potential barrier height $\Delta E$ can be quantitatively derived from temperature dependent mobility data using
\begin{equation}
\mu(T)=\mu_0 \exp \left( \Delta E/k_B T\right)
\label{muT}
\end{equation} 
where the mobility prefactor $\mu_0=\nu_0 T^{-1/2}$ with $\nu_0$ being temperature independent. The behavior results from thermionic emission~\cite{seto1975} .

We observed two different slopes for the full temperature range and attribute it to different transport regimes. Usually, the high temperature regime is assigned to thermionic emission over the barrier, whereas at low temperature thermally assisted tunnelling is dominant~\cite{orton1980}. We evaluated the activation energy in the high temperature regime,  enabling us to relate it to the other parameters, such as the conductivity, charge carrier density, and mobility, derived at 300~K. 

Figure~\ref{fig:deltaE} shows the activation energy $\Delta E$ $vs.$ carrier concentration in the high temperature regime. For the undoped material, we observe a monotonously decreasing mobility with increasing carrier density, as expected. Doping also results in a decrease in $\Delta~E$ up to a doping level of 2~at$\%$, in the case of $V_g$=0~V as well as $V_g$=20~V, indicating a doping induced lowering of the grain boundary energy barriers. At the high doping level of 10~at$\%$, however, a higher potential barrier of 95~meV and 113~meV, repectively, is present, which we ascribe to the influence of structural disorder.

\section{Discussion}

Our experimental results indicate that extrinsic doping influences the electrical parameters of the nanocrystalline ZnO in a complex way. In the following, we discuss them in a context of charge carrier density variation superimposed by crystallite size modification and structural disorder.
 
 The XRD spectra show a significant decrease in the (002) peak height, which can be accounted for by the additional disorder and transistion to the amorphous phase, induced by the high percentage of dopants in the material. 
Analyzations of the spectra with the Debye-Scherrer method indicate that the crystallite size decreases with increasing doping level, resulting in a higher grain boundary density. In contrast, the maximum particle size, as seen with SEM, is increasing up to a doping level of 2~at$\%$. We note that the discrepancy between crystallite sizes estimated by the Debye-Scherrer method and the particle sizes seen in SEM micrographs has been previously reported for sputtered ZnO:Al layers~\cite{wanderka1998} and for other nanocrystalline systems.~\cite{devi2007} The origin of this difference is attributed to particles, consisting of a few crystallographic domains, i.e., crystallites. Between those grains the grain barrier is being formed. One also has to keep in mind the possiblity, that the ordering within the first few nanometers of the material can be different from the bulk material.

The charge carrier density in the nominally undoped samples, i.e., denoted as  0~at$\%$ doping level, is surprisingly high in our case. Actually, considering the bulk energy gap around 3.3 eV,~\cite{srikant1998} one would expect an undoped system to be insulating. However, our experimentally observed charge carrier densities in the undoped case are in good agreement with literature for sputtered ZnO~\cite{martins2005} and sol--gel processed ZnO.\cite{lin2008} The unintentional intrinsic doping of zinc oxide is probably due to oxygen and zinc lattice defects,~\cite{look1999} as well as due to hydrogen interstitials, identified in crystalline systems~\cite{vandewalle2000, hofmann2002}. Also, for a different sol--gel route, synthesis educts in nc-ZnO:Al have been found by~\cite{orlinskii2008}. These impurities can also act as dopants.

Depending on the Al doping level, the charge carrier density increases from 8.9$\cdot$10$^{17}$~cm$^{-3}$ in the nominally undoped case to 5.6$\cdot$10$^{18}$~cm$^{-3}$ at 10~at$\%$, as expected.
We note that in some cases the opposite doping dependence on the charge carrier density can be observed in Hall measurements~\cite{paul2002}. There, a lowered $n$ at high doping levels was reported. This trend could not be confirmed by our measurements.

For low doping levels, the growing free charge carrier density leads to an increase in conductivity, whereas at high doping levels, the induced disorder lowers the mobility and thus counteracts the positive effect of a doping induced high carrier concentration on the conductivity. Therefore, we find a maximum in $\sigma$ at 0.8~at$\%$ extrinsic doping level (Figure \ref{fig:mucondfitn}). We note that the values of conductivity and mobility in zinc oxide nanoparticles studied in our work are low in comparison to other publications~\cite{lin2008,sagar2005}. We attribute this effect to the use of thin nc-ZnO layers, which are required by the field effect transistor geometry. An increase of the conductivity and the mobility proportional to the layer thickness has been reported previously.~\cite{schuler1999, alam2001} Also, a contact limitation can not be neglected for the determination of $\sigma$, even though we expect this effect not to be dominant. Assuming a very high contact resistance of 10~M$\Omega$m leads to a conductivity and free charge carrier increment of only a factor of two.

At $V_g$=0~V we observe a slightly higher mobility at 0.8~at$\%$ as compared to the nominally undoped sample, which we attribute to a doping induced potential barrier height lowering typically found in polycrystalline systems,~\cite{orton1980, seto1975, baccarani1978, kimura2001} which was also reported for sol--gel synthesised nc-ZnO:Al~\cite{sagar2005, paul2002}.
Further extrinsic doping of the nc-ZnO with Al leads to a decreasing electron mobility with increasing doping level, as shown in Figure~\ref{fig:mucondfitn}. At $V_g$=20~V, we observe decreasing mobilities for increasing doping levels. Here, the charge carrier mobility is thus proportional to the crystallite size (see Table~\ref{dbs}).

In order to investigate the charge transport properties in more detail, we performed temperature dependent mobility measurements.  As seen in Figure~\ref{fig:mut0at} for the nominally undoped sample, two different transport regimes are indicated by the changing slope of the temperature dependent electron mobility. This effect is well known for polycrystalline semiconductors; the high temperature regime is usually assigned to thermionic emission over energy barriers at the grain boundaries, whereas at low temperature thermally assisted tunnelling~\cite{orton1980} or variable range hopping between donor states within the crystallites~\cite{natsume1992} is dominant. The activation energy which we derived in the high temperature regime initially decreases with increasing doping level, with a minimum barrier height at 2~at$\%$. The sample with the highest doping shows an increase in the activation energy, as seen in Figure~\ref{fig:deltaE}. The decreasing activation energy for increasing doping level up to 2~at$\%$ is consistent with the above-mentioned doping induced barrier lowering. However, as the electron mobility in our systems mostly does not show the expected increase with lowered energy barrier, the charge transport is not limited by thermionic emission over the grain boundaries. Note that high doping levels in crystalline systems also lead to a diminished temperature dependence due to the reduction of the mean free path within the crystallites. 

Two possible scenarious concerning the decreasing charge carrier mobility due to the extrinsic doping, (i) intercrystallite scattering due to a higher density of grain boundaries, and (ii) scattering within the crystallites at the defects in the band gap, could both explain our experiments by a formation of additional scattering centers. We considered the first option following the approach of Hossain et al.~\cite{hossain2003} However, the increasing number of grain boundaries shows a lower influence on the mobility as compared to the lowered barrier height, which is why we discard this explanation: the calculated mobilities do not show the experimentally observed decrease with raised extrinsic doping level. The second scenario, scattering by disorder and impurities due to the high extrinsic doping levels and the sol--gel processing, becomes thus the most probable option governing the charge carrier mobility above 0.8~at$\%$ doping. The microscopic origin of these additional trap states is unclear as of yet; an ionic, electronic and/or structural origin of the scattering centers is possible. 

In contrast to the effect of extrinsic doping, the mobility steadily increased due to accumulation of charge carriers in the field effect transistor channel (see Figure~\ref{fig:accdope}), whereas the trap density remained unchanged.The activation energy, shown in Figure~\ref{fig:deltaE}, decreases correspondingly. Thus, in the case of accumulation, where the density of scattering centers remains constant, the experimental evidence is in favour of carrier concentration induced barrier height lowering governing the charge transport properties.

\section{Conclusions}
We synthesized Al doped zinc oxide nanoparticles via a sol--gel route, resulting in average crystallite sizes between 20~nm to 7~nm corresponding to doping concentrations from 0~at$\%$ to 10~at$\%$, respectively. The synthesized nanocrystals possess the Wurtzite structure as derived from XRD and SEM studies. We investigated the charge transport by using field effect transistor structures on SiO$_2$ with Au bottom contacts. The accumulation of charge carriers in thin films of undoped zinc oxide nanoparticles results in higher charge carrier mobilities. In contrast, despite the extrinsic Al doping leading to an increasing charge carrier density as well, the mobility decreases at higher doping levels. This is attributed to the formation of additional trap states of ionic, electronic and/or structural origin acting as scattering centers. High doping levels lead to smaller potential barrier heights at the nanoparticle interfaces. The potential barrier height derived from temperature dependent measurements reaches its minimum at a doping concentration of 2~at$\%$. Further doping, however, leads to an increase of the charge transport activation energy, as the doping-induced structural defects become dominant. The maximum conductivity was achieved at the doping concentration of 0.8~at$\%$ at room temperature. The interplay between charge carrier density, barrier height and grain boundary density leads to a lowering of the mobility with increasing doping level. To conclude, we demonstrate that the Al doping of zinc oxide nanoparticles by a sol--gel process is adjustable and can be controlled, but it does not automatically lead to improved electric transport properties as it simultaneously generates structural disorder, thus increasing the scattering of charge carriers.

\section*{Acknowledgements}
V.D.'s work at the ZAE Bayern is financed by the Bavarian Ministry of Economic Affairs, Infrastructure, Transport and Technology.

\newpage 


\section*{References}
\begin{thebibliography}{10}

\bibitem{orlinskii2008}
Orlinskii S B, Schmidt J, Baranov P~G, Lorrmann V, Riedel I, Rauh D
  and Dyakonov V  2008 
{\em Phys. Rev. B} {\bf 77} 115334.


\bibitem{xue2006}
Xue S W, Zu X T, Zheng W G, Deng H X and Xiang X 2006 
{\em Physica B: Condensed Matter} {\bf 381} 209


\bibitem{lin2008}
Lin J P  and Wu J M 2008
{\em Appl. Phys. Lett.} {\bf 92} 134103

\bibitem{paul2002}
Paul G K, Bandyopadhyay S and Sen S K 2002 
{\em Phys. Stat. Sol. A} {\bf 191} 509

\bibitem{sagar2005}
Sagar P, Kumar M and Mehra R M 2005  
{\em Materials Science--Poland} {\bf 23} 685


\bibitem{schuler1999}
Schuler T and Aegerter M A 1999  
{\em Thin Solid Films} {\bf 351} 125--131

\bibitem{alam2001}
Alam M J and Cameron D C 2001
{\em The 47th international symposium: Vacuum, thin films, surfaces/interfaces, and processing NAN06} {\bf 19} 1642


\bibitem{cheng2007}
Cheng H C, Chen C F and Tsay C Y 2007  
{\em Appl. Phys. Lett.} {\bf 90} 012113

\bibitem{beek2004}
Beek J E, Wienk M M and Janssen R A J 2004,  
{\em Adv. Funct. Mater.} {\bf 16} 1009

\bibitem{gilot2007a}
Gilot J, Wienk M M, and Janssen R A J 2007  
{\em Appl. Phys. Lett.} {\bf 90} 143512

\bibitem{gilot2007b}
Gilot J, Barbu I, Wienk M M and Janssen R A J 2007,  
{\em Appl. Phys. Lett.} {\bf 91} 113520

\bibitem{hadipour2008}
Hadipour A, deBoer B and Blom P W M 2008
{\em Adv. Funct. Mater.} {\bf 18} 169

\bibitem{koster2007}
Koster L J~A, van~Strien W~J, Beek W J~E and Blom P W~M 2007 
{\em Adv. Funct. Mater.} {\bf 17} 1297--1302

\bibitem{langton1958}
Langton N H and Matthews D 1958
{\em British J. Appl. Phys} {\bf 9} 453

\bibitem{marien1976}
Marien J 1976 
{\em Phys. Stat. Sol. A} {\bf 38} 513

\bibitem{srikant1998}
Srikant V and Clarke D R 1998 
{\em J. Appl. Phys} {\bf 83} 5447

\bibitem{zhong2004}
Zhong L W 2004  
{\em J. Phys. Condens. Matter} {\bf 16} R829

\bibitem{meulenkamp1998}
Meulenkamp E 1998 
{\em J. Phys. Chem. B} {\bf 102} 5566

\bibitem{ellmer2008}
Ellmer K and Mientus R 2008
{\em Thin Solid Films} {\bf 516} 4620

\bibitem{orton1980}
Orton J W, and Powell M J 1980  
{\em Rep. Prog. Phys.} {\bf 43} 1263


\bibitem{seto1975}
Seto J Y~W 1975
{\em J. Appl. Phys} {\bf 46} 5247

\bibitem{baccarani1978}
Baccarani G, Ricco B and Spadini G 1978 
{\em J. Appl. Phys} {\bf 49} 5565

\bibitem{hossain2003}
Hossain F M, Nishii J, Takagi S, Sugihara T, Ohtomo A, Fukumura T 
Koinuma H, Ohno H, and Kawasaki M 2003,
{\em J. Appl. Phys} {\bf 94} 7786

\bibitem{jones1938}
Jones F W 1938  
{\em Proceedings of the Royal Society of London. Series A, Mathematical and
  Physical Sciences} {\bf 166(924)} 16--43

\bibitem{kimura2001}
Kimura M, Inoue S, Shimoda T and Sameshima T 2001 
{\em Japan. J. Appl. Phys} {\bf 40} 5237

\bibitem{shur1989}
	Shur M, Hack M, and Shaw J G 1989
	{\em J. Appl. Phys} {\bf 66} 3371

\bibitem{wanderka1998}
Sieber I, Wanderka N, Urban I,  D\"orfel I,  Schierhorn E, Fenske F and Fuhs W 1998  
{\em Thin Solid Films} {\bf 330} 108

\bibitem{devi2007}
Devi R, Purkayastha P, Kalita P K and Sarma B K 2007
{\em Bull. Mater. Sci.} {\bf 30} 123


\bibitem{martins2005}
Martins R, Barquinha P, Pimentel A, Pereira L and Fortunato E 2005 
{\em Phys. Stat. Sol. A} {\bf 202} R95

\bibitem{look1999}
Look D C, Hemsky J W and Sizelove J R 1999  
{\em Phys. Rev. Lett.} {\bf 82(12)} 2552

\bibitem{vandewalle2000}
Van de Walle C G 2000  
{\em Phys. Rev. Lett.} {\bf 85} 1012

\bibitem{hofmann2002}
Hofmann D M, Hofstaetter A, Leiter F, Zhou H, Henecker F, Meyer B K, Orlinskii S B, Schmidt J and  Baranov P G 2002
{\em Phys. Rev. Lett.} {\bf 88} 045504

\bibitem{natsume1992}
Natsume Y, Sakata H, Hirayama T and Yanagida H 1992
{\em J. Appl. Phys} {\bf 72} 4203






\end{thebibliography}
\end{document}